\documentclass[runningheads,a4paper]{llncs}
\pdfoutput=1
\usepackage{hyperref}
\usepackage{pdfpages}
\usepackage{esvect}
\usepackage{amssymb}
\usepackage{graphicx}
\usepackage{url}
\urldef{\mailsa}\path|{bibek.behera, manoj.joshi}@searshc.com| 
\urldef{\mailsb}\path|{Abhilash.KK, Ansari.MohammedIsmail}@searshc.com|
\newcommand{\keywords}[1]{\par\addvspace\baselineskip
\noindent\keywordname\enspace\ignorespaces#1}

\begin{document}

\mainmatter  

\title{Distributed Vector Representation Of Shopping Items, The Customer And Shopping Cart To Build A Three Fold Recommendation System.}

\titlerunning{Lecture Notes in Computer Science: Authors' Instructions}

%
%
\author{Bibek Behera, Manoj Joshi, Abhilash KK, Mohammad Ansari Ismail}
%

\institute{Sears Holding Corporation\\
\mailsa\\
\mailsb\\
}

%
%

\toctitle{Lecture Notes in Computer Science}
\tocauthor{Authors' Instructions}
\maketitle

\begin{abstract}
The main idea of this paper is to represent shopping items through vectors because these vectors act as the base for building embeddings for customers and shopping carts. Also, these vectors are input to the mathematical models that act as either a recommendation engine or help in targeting potential customers. We have used exponential family embeddings as the tool to construct two basic vectors - product embeddings and context vectors. Using the basic vectors, we build combined embeddings, trip embeddings and customer embeddings. Combined embeddings mix linguistic properties of product names with their shopping patterns. The customer embeddings establish an understanding of the buying pattern of customers in a group and help in building customer profile. For example a  customer profile can represent customers frequently buying pet-food. Identifying such profiles can help us bring out offers and discounts. Similarly, trip embeddings are used to build  trip profiles. People happen to buy similar set of products in a trip and hence their trip embeddings can be used to predict the next product they would like to buy. This is a novel technique and the first of its kind to make recommendation using product, trip and customer embeddings.

\keywords{Product Embedding, shopping cart, vector representation }
\end{abstract}

\section{Introduction}

Word embeddings have changed the research world in natural language processing. The word to vector model proposed by Mikolov et al. \cite{mikolov2013distributed} took the field of AI and NLP by storm. Later, word2vec model was extended to sentences and documents as sent2vec and doc2vec models \cite{le2014distributed} and found their ways into myriad of applications. 

In recent times, there has been research to bring similar concepts into other domains. Rudoplh et al. \cite{rudolph2016exponential} has shown how to apply word2vec model in various domains like movies, grocery and medical science using the concept of family embeddings. In case of grocery, we obtain embeddings for each product by iterating over transactions that is available in huge quantities from the transaction logs of retail companies. Product embedding is the manifold space of products where their abstract representation makes semantic sense. For example \textit{cat-food} and \textit{cat-accessories} are nearby in the product embedding space. Secondly the vector difference between \textit{cat-food} and \textit{cat-accessories} is same as the vector difference between \textit{dog-food} and \textit{dog-accessories}. These kind of semantic structure hidden in the product embeddings speak of the quality of the representations and lead us to build the trip and customer embeddings on the foundation of product embeddings 

In this paper we begin with the mathematical model called exponential family embedding (Ef-emb) \cite{rudolph2016exponential} and the characteristics of product embeddings and context vectors. Then we construct a recommendation system that finds similar and co-occurring items. Here similar items mean items that can replace each other. For example \textit{"Beverage A"} can be a replacement of \textit{"Beverage B"}. By co-occurring products, we mean products that can be bought together. 


We then deploy a sent2vec model on the labels of products. This gives the product a linguistic prospective. We combine the sentence embeddings with the product embeddings and obtain combined embeddings. We investigate the properties of all three embeddings. 

Then we extend the concept of product embedding to trips and customers. A trip consists of all the products bought by a customer in one single transaction. The motivation is that we find cluster amongst trips so that we can recommend products to customers based on their current cart. Similarly we find customer embeddings and then find meaningful customer clusters using K-means algorithm based on Lloyd's algorithm \cite{lloyd1982least}. These groupings lead us to  departments of products that they buy frequently. The frequently observed departments make the profile of the customer group. For example a profile of customers can consist of \textit{girl tops}, \textit{preschool} and \textit{tod boy sportswear} probably because the customers represent parents whose kids are in the age range of 3 to 5. Having done such profiling, offers and discount can be targeted to a customer if she happens to be similar to a particular group based on the profile of the group.

The road map for the rest of the paper is as follows. First we lay down the specifics of the theory behind family embeddings and their extension from word2vec model but with their own peculiarities in section \ref{sect:Efe}. Then we demonstrate the recommendation engine in section \ref{sect:re}. Henceforth, we show a sent2vec model and a combined model in section \ref{sect:s2v}. Afterwards, we explain the process behind construction of trip embeddings and analysis of trip in section \ref{sec:te} followed by similar construction of customer embeddings and customer profiling in section \ref{sec:ce}. The description of entire pipeline is explained in section \ref{sec:sr}. This is followed by conclusion and future work in section \ref{sec:con}. 

\section{Exponential family embeddings}
\label{sect:Efe}
Exponential family embeddings (Ef-emb) \cite{brown1986fundamentals} are statistical tools that generalise the technique to capture contextual information for various data sets and their varying distributions. They have been employed by Rudolph et al. \cite{rudolph2016exponential} to extend the word2vec model to different domains like grocery, movies, \textit{etc}. This model requires two inputs -the data point and the context. For example, in grocery, the item bought is the data point while context is all other items bought in that cart or trip. In case of movies, the movie being watched is the data point while the context is all the movies rated by the same person. Thus Ef-emb gives the user the flexibility to choose their context function. 

The motivation behind using an Ef-emb is to derive useful features of the data. Context vector, distribution of the data points and their embeddings are the key ingredients that are part of a loss function which has to be minimized iteratively. By distribution of data points, we mean to say the nature of data points can be continuous or discrete values. For example in grocery data we can have product quantities which are numerical in nature hence we can use Bernoulli or Poisson distribution \cite{mccullagh1984generalized}. In case of word2vec the words are binary in nature hence we can use Bernoulli distribution. In case of neural activity, their time of activity is a continuous variable wherein we use the Gaussian distribution.  Thus, Ef-emb can effectively consume any kind of data and produce distribution of vector representation of data points. We also get to know the semantic structure e.g. movies of the same genre are clubbed together or some neurons have same time of activity even if they are spatially distant. 

\subsection{Similar and co-occuring products}
\label{sub:scp}
In case of shopping data, the motivation is to get similar and co-occurring products \cite{linden2003amazon}. According to the concept described in Rudolph et. al 2016 \cite{rudolph2016exponential}, product embeddings can help in determining similar products by using cosine distance in the vector space to generate nearest neighbors as shown in equation \ref{eq:sim} where \( \rho \) stands for product embeddings. In other words if the product embeddings of a pair of products have high cosine similarity, then they can replace each other.
\begin{equation}\label{eq:sim}
similarity\_score = cosine\_distance\left(\overrightarrow{\rho_x}, \overrightarrow{\rho_y} \right)
\end{equation}
Co-occurring items are obtained by calculating inner product between all pairs of product embeddings and context vectors. It has been found if a pair of product co-occur the inner product of product embedding of x (\( \rho_x \)) and context vector of y (\( \alpha_y \)), as shown in equation \ref{eq:cooc_score} is higher. If they do not co-occur their inner products tend to be negative. 
\begin{equation} \label{eq:cooc_score}
cooccurrence\_score = cosine\_distance\left(\overrightarrow{\rho_x}, \overrightarrow{\alpha_y} \right)
\end{equation}
Using the two models for similar and co-occuring products we build a basic recommendation engine. Some of the examples from the recommendation system has been shown in Table \ref{table:rec}.

\begin{table}[h!]

\begin{tabular}{||c |c| c| c||} 
 \hline
 Product name & rec product 1 & rec product 2 & rec product 3 \\  
 \hline\hline
 Coke & Mt Dew & Sprite & Pepsi \\ 
 Snickers  & Twix Caramale & Kit Kat & M n M Peanut \\
 Aquafina water & Dasani water & Smart sense water & Pepsi Cola \\
 Angel soft Tissue & Scott Bath Tissue & Scott Towel & Paper Towels \\
 Lay's sour & Doritos Nacho & Lay's classic potato & Cheetos \\ 
 \hline
\end{tabular}
\caption{Some examples from the recommendation engine}
\label{table:rec}
\end{table}

\section{Recommendation engine}
\label{sect:re}
We have used six month's transaction data of Sears Holdings Corporation. Firstly, we found all products that were bought in a single transaction henceforth called a trip. We consider those trips with 5 or more products, so that we have a context for each product. Each product is a data point and the products bought in the transaction become the context word. Now this data is fed to the Ef-emb model which assume a Bernoulli distribution since we just account for the presence or absence of each product. We employ the algorithm implemented by Rudolph et. al 2016 \cite{rudolph2016exponential}. The algorithm runs for 1000 iterations and the data has approximately 20 million trips and 500 thousand products. After completion, it creates two arrays  - \( \rho \) for product embeddings  and \( \alpha \) for context vector for each product. Each vector has a length of 100. These vectors are fed to Annoy which is a fast c++ library with python bindings. Annoy generates an approximate nearest neighbour model as proposed by Arya et. al 1998 \cite{arya1998optimal} for a million products . This model can be employed in a realtime recommendation system \cite{schafer1999recommender}. To find similar items, we provide product embedding and labels. Given a product id of the reference item, the model generates similar items as per equation \ref{eq:sim}. We find co-occuring items as discussed in section \ref{sub:scp}.

\section{Visualisation of product embedding and context vectors}
We used TSNE to visualise the 100 dimension vector. TSNE is available as a python or R package and was developed by Maaten et. al. 2008 \cite{maaten2008visualizing}. It converts n-dimensional vector space to m-dimensional vector space using a probabilistic approach and minimizes the KL divergence between the 2 data sets. The TSNE algorithm works well for data that is difficult to classify at higher dimension. The beauty of TSNE is that even after converting to lower dimension it retains the clusters in the higher dimensions to a high degree of precision. So we can interpret patterns in data by actually visualising them in 2D or 3D.  In reality we see a multiple number of interpretable clusters and find relation easier to interpret. For example, \textit{cat-food} and \textit{cat-accessories} are placed closeby. Similarly \textit{dog-food} and \textit{cat-food} are clumped nearby. We also see semantic relation as show in \ref{eq:vec_diff}
\begin{equation} \label{eq:vec_diff}
\vv{cat food} - \vv{cat accessories} = \vv{dog food} - \vv{dog accessories}
\end{equation}
In the 2D visualisation we found cluster that represents \textit{beverage} like \textit{coke}, \textit{pepsi}, \textit{dew}, etc and also the cluster of \textit{water products} is nearer to that of \textit{beverage}. 

\begin{figure}
    \centering
    \includegraphics[width=1\textwidth]{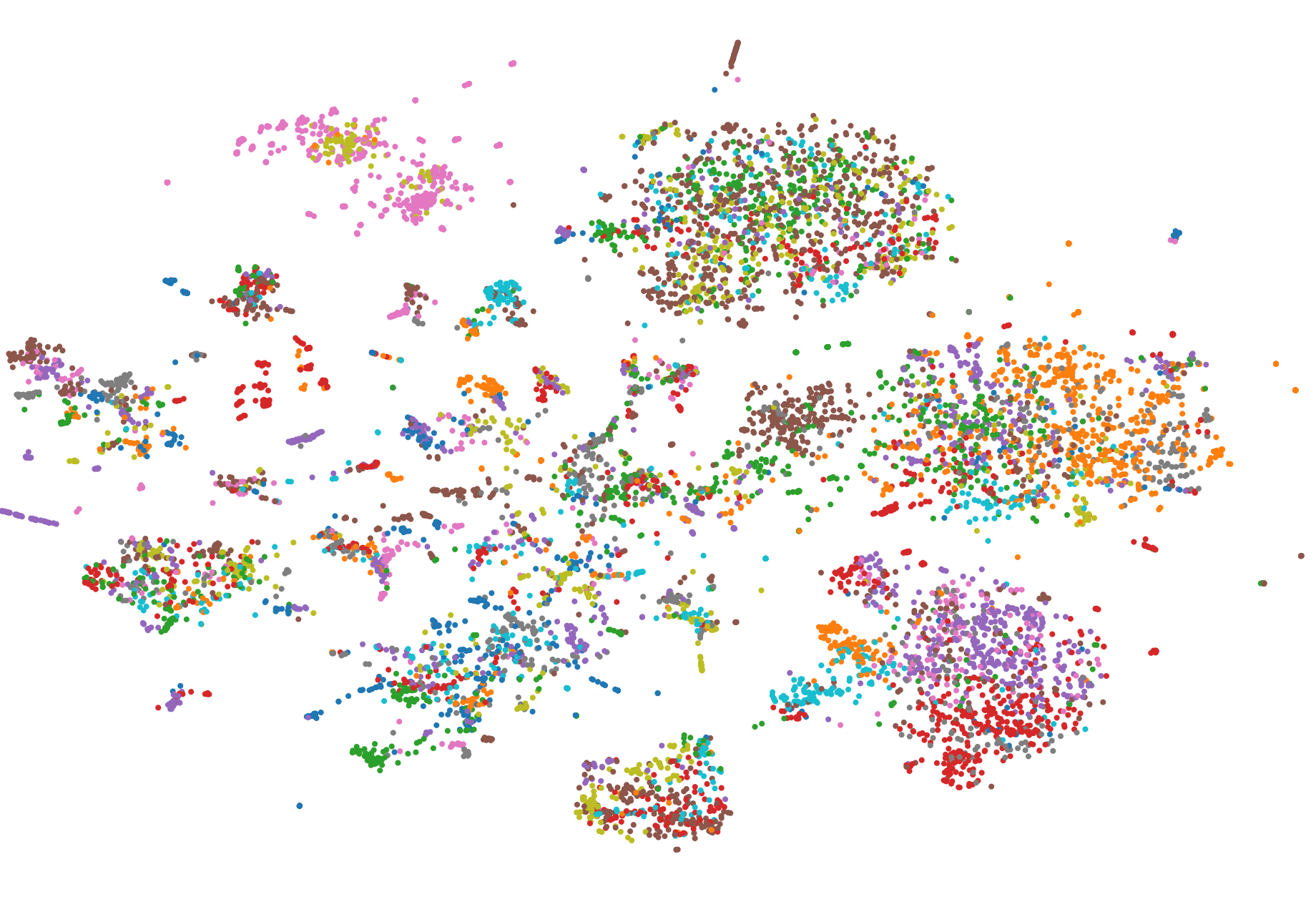}
    \caption{Product embedding after applying TSNE.}
  
    \label{fig:we}
\end{figure}

For both vectors, we ran TSNE for 1000 iteration and converted a 100 dimension vector space to 2 dimension vector space. The visualisation of product embedding is shown in figure \ref{fig:we}. We verify the cluster mathematically by running the recommendation system. The results returned by recommendation system which works on the 100 dimensional vector concurred with the 2D representation using TSNE.

\section{Sent2Vec model and combined model}
\label{sect:s2v}
The motivation is to find linguistic patterns in products and then combine them with product embeddings to form a vector representation that is an amalgamation of linguistic and shopping patterns of products.

We generate the labels of around five hundred thousand products and give it as input to the sent2vec model. The sent2vec model then creates vectors based on the context of every word present in those product names. The characteristic of this model is that it finds clusters of products bearing similar names while the characteristic of product embeddings is that they are capable of finding clusters of products that are conceptually similar. To bridge this gap between the two models, we build a combined model that takes vector representation of product embeddings and sentence embeddings and concatenate them. 

We use TSNE to project them to a 2D map. In the projection of sentence embeddings, \textit{Swimwear} and \textit{Beachwear} appear far apart but in the projection of combined embeddings they appear close. Another example is that of the \textit{cat food} and \textit{dog food} that are distant in the sent2vec model come closer in the combined model. This is probably due to the properties that are brought by the product embeddings. Although combined model is merely a merger of sentence embeddings and product embeddings but some examples show that the properties of product embeddings seem to improve the quality of clusters created in sent2vec model.

The combination of sentence embeddings and product embeddings space can be seen as a merger of natural language vectors and shopping data patterns. Thus by combining these spaces together we have embeddings that are more useful conceptually. We have not verified mathematically whether the combined model is an improvement over product embeddings. That restrains us from using the combined model for recommendation.

\section{Application of product embedding}
Having obtained the product embeddings of all products, we use them to find out patterns in trips and customers. By patterns we mean any information which could help us profile the trips and customers. By profiling we mean to summarise the products that are being bought by a group of trips or a cluster of customers. This information is particularly helpful in targetting customers for offers and discounts. The only information we have prior to obtaining product embeddings are the departments of the product and we will use these departments to profile a cluster of trips and customers. Everytime a new product appears in a department, we could find a potential customer and then target her by giving offers and discounts in that department category. Some of the department categories have been listed in table \ref{tab:tags}.

\begin{table}[h!]

\begin{tabular}{||c |c| c| c||} 
 \hline
 Carbonated soft drink & sports toys & arts n crafts & salty snacks \\  
 Halloween candy & Juices and drinks & Paper supplies  & Bath tissue \\ 
 Oral care  & Towels & Cups and plates & Dog treats \\
 Cat food & Mens underwear & Cat foods & Household Chemicals \\
 Mens Sports Socks & Hair Care & Antifreeze solvents & Beach Towels \\
 
 \hline
\end{tabular}
\caption{Some frequent department labels}
\label{tab:tags}
\end{table}

\section{Trip Embedding}
\label{sec:te}
The idea is to convert each trip into vector by taking mean of the sum of  the product embeddings \( \rho \) of all the \textit{n} products bought in one single trip as shown in equation \ref{eq:trip}. 
\begin{equation} \label{eq:trip}
trip\_embedding[i] = \frac{1}{n}\sum_{i=1}^{n} \rho[i]
\end{equation}

\begin{figure}
    
  \centering
    \includegraphics[width=0.8\textwidth]{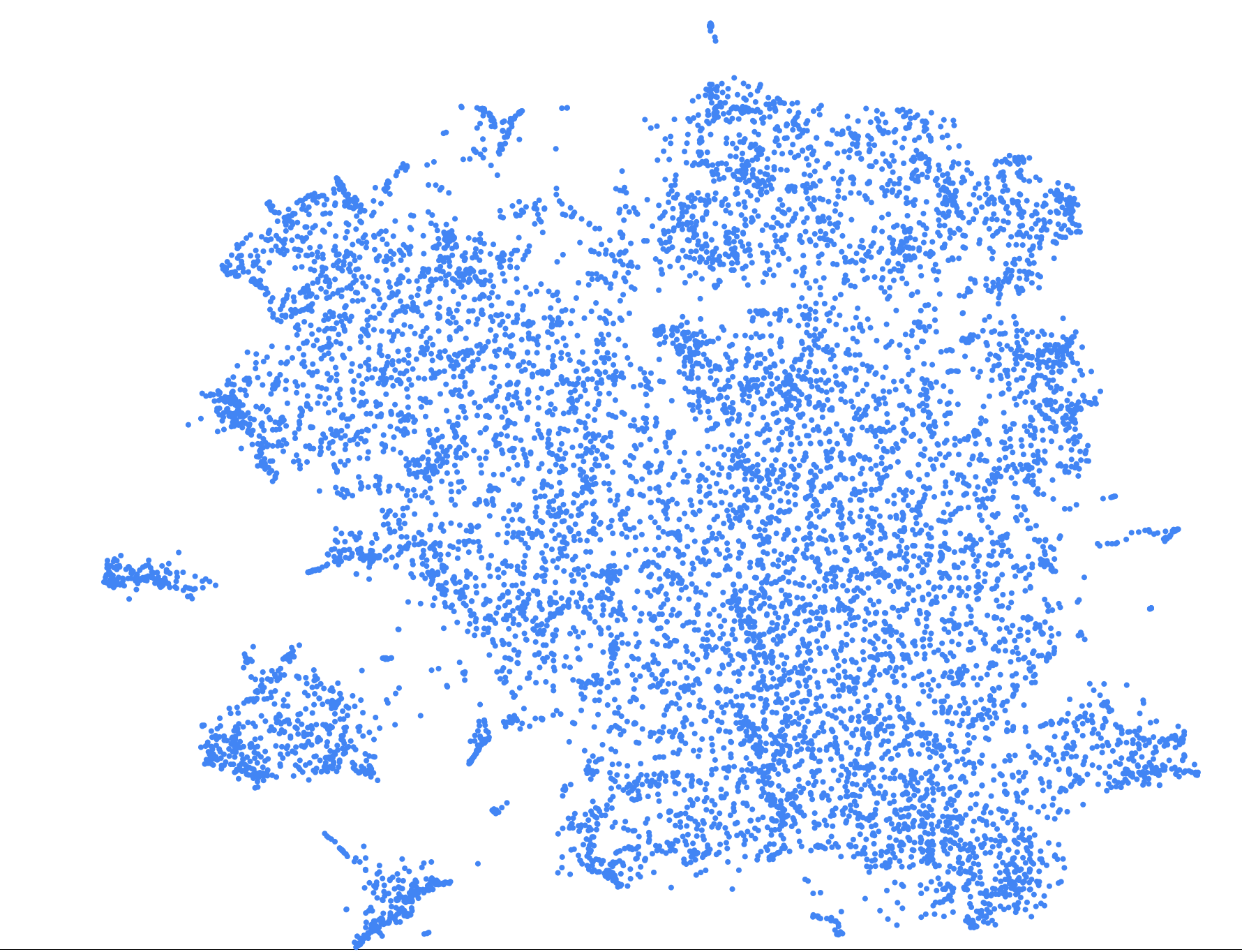}
    \caption{Trip embedding after applying TSNE.}
    \label{fig:te}
    
\end{figure}

In all there were 22 million trips and we converted them to vectors. We took a sub sample of 10k vectors after filtering all trips that had 5 or more products. This was done because we needed to extract buying patterns in trips. If the number of products in a trip are lesser than a threshold (here 5) then we may not get a significant pattern. After employing TSNE we project them onto a 2d map where we could actually see clusters as shown in figure \ref{fig:te}.

\subsection{Trip analysis}
After obtaining the projection onto 2d map, we ran the KNN over the trip embedding space and extracted 5 clusters. The noticeable point is that despite having no common items some trip embeddings were paired together. We call such trip fake pairs. To estimate number of fake pairs,  we find nearest neighbours using Annoy and then if they do not have similar items we call them fake pairs. For each cluster we made the trip analysis  as shown in table \ref{tab:ta}.
\begin{table}[h!]
\centering
\begin{tabular}{||c |c| c| c||} 
 \hline
Cluster Id & True pairs & Fake pairs & Cluster score \\  
 \hline
1 & 95144 & 210946 & 0.310836682022 \\
2 & 30176 & 83674  & 0.265050505051 \\
3 & 35666 & 201484 & 0.150394265233 \\
4 & 36262 & 101978 & 0.262311921296 \\
5 & 14256 & 90504  & 0.136082474227 \\
 \hline
\end{tabular}
\caption{Trip Analysis}
\label{tab:ta}
\end{table}

\subsection{Trip Profiling}
The second analysis was to profile each cluster based on the frequency of departmental tags assosciated with product bought in the trips. The cluster and their top tags have been shown in table \ref{tab:tp}.

\begin{table}[h!]

\begin{tabular}{||c |c| c| c||} 
 \hline
 Cluster size & Dept 1 & Dept 2 & Dept 3 \\  
 \hline
 3401 & HOUSEHOLD CHEMICALS & FABRIC CARE SUPPLIES & TAKE HOME \\
 1265 & GIRL TOPS  & BOY KNIT TOPS & GRL FASH BOTTOMS \\
2635 & PERSONAL HYGIENE & ORAL CARE & TAKE HOME \\
1536 & GIRL TOPS  & PRESCHOOL & TOD BOY SPORTSWEAR \\
1164 & ROUTE 66 & BE TOPS & JUNIORS \\
 
 \hline
\end{tabular}
\caption{Trip Profile}
\label{tab:tp}
\end{table}

\section{Customer Embedding}
\label{sec:ce}
The idea is to represent each customer through a vector by averaging the product embeddings of all the \textit{n} products bought in all trips over a period of six months by a customer \textit{i} as shown in equation \ref{eq:trip}. 
\begin{equation} \label{eq:trip}
customer\_embedding[i] = \frac{1}{n}\sum_{j=1}^{n} \rho[i]
\end{equation}

\begin{figure}
    \centering
    \includegraphics[width=0.8\textwidth]{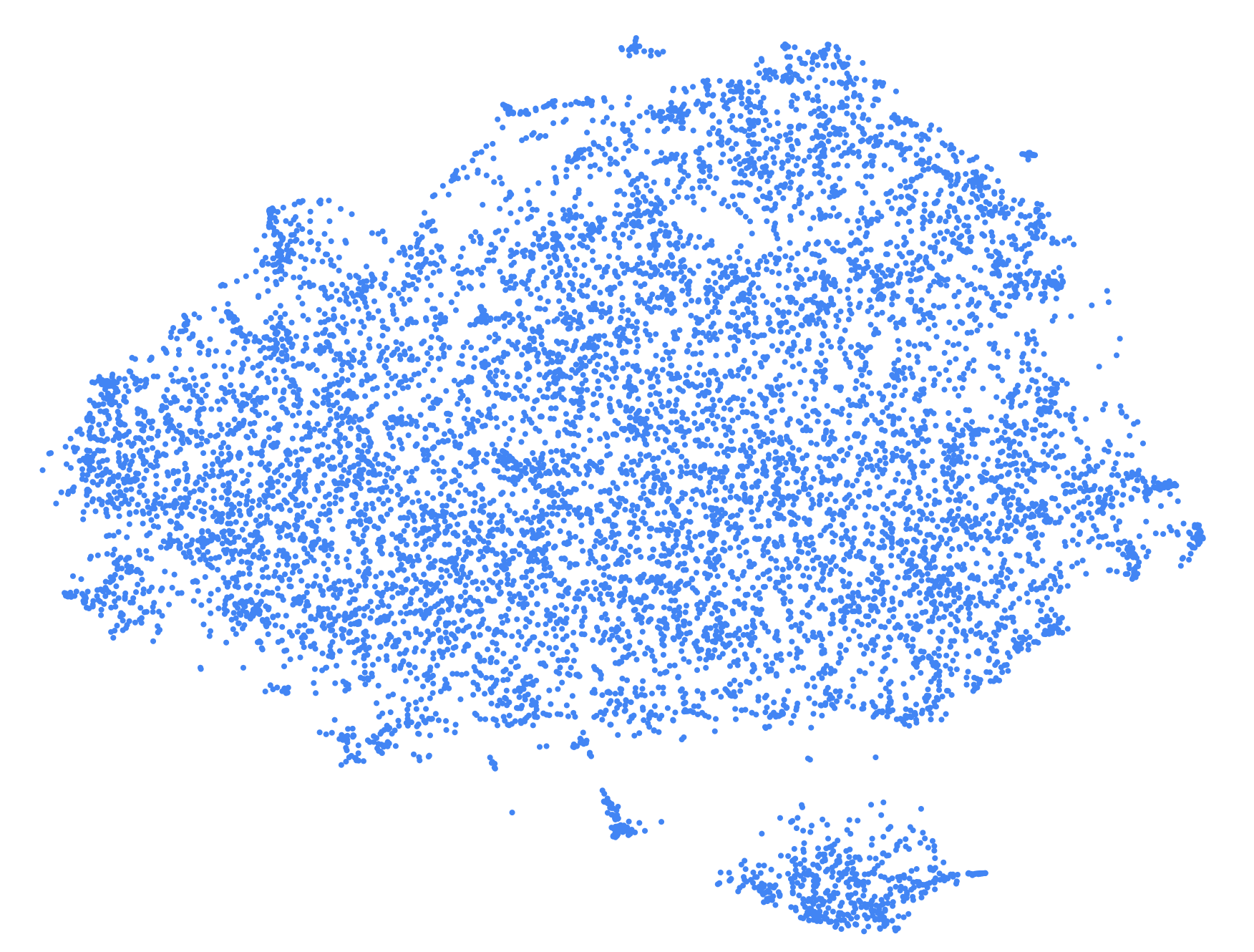}
    \caption{Cluster embedding after applying TSNE.}
  
    \label{fig:ce}
\end{figure}

In all, there were 16 million trips and we converted them to vectors. We took a subsample of 10k vectors after filtering all trips that had 5 or more products. This was done because we needed to extract cluster of customers. In smaller trip, appearance of a pattern is not significant. After employing TSNE we project them onto a 2d map where we could actually see clusters as shown in figure \ref{fig:ce}.

\subsection{Customer analysis}
After obtaining the projection onto 2d map, we ran the KNN over the customer embedding space and extracted 5 clusters. The customer analysis is similar to trip analysis  as shown in table \ref{tab:ca}.
\begin{table}[h!]
\centering
\begin{tabular}{||c |c| c| c||} 
 \hline
Cluster Id & True pairs & Fake pairs & Cluster score \\  
 \hline
 1 & 50526 & 74934 & 0.402725968436 \\
2 & 40154 & 77926 & 0.340057588076 \\
3 & 188692 & 123338 & 0.604723904753 \\
4 & 31328 & 5752 & 0.844875943905 \\
5 & 151638 & 155802 & 0.493227946916 \\
 
 \hline
\end{tabular}
\caption{Customer Analysis}
\label{tab:ca}
\end{table}

\subsection{Customer Profiling}
Customer profiling was done in the same manner as trip profiling and has been shown in table \ref{tab:cp}. As shown in customer profiles, we can identify that cluster 1 reflects sportwear, cluster 2 is about tops, cluster 3 about chemicals, cluster 4 has more items related to hygiene and cluster 5 is about beverages. This way we can find clusters and give them profiles. 

\begin{table}[h!]

\begin{tabular}{||c |c| c| c||} 
 \hline
 Cluster size & Dept 1 & Dept 2 & Dept 3 \\  
 \hline
 1394 & GIRL TOPS & BOY KNIT TOPS & TOD BOY SPORTSWEAR \\
1312 & ROUTE 66 & BE KNIT TOPS & BE TOPS \\
3467 & TAKE HOME & HOUSEHOLD CHEMICALS & NEW AGE/ WATER \\
412 & HOUSEHOLD CHEMICALS & PERSONAL HYGIENE & FABRIC CARE SUPPLIES \\
3416 & TAKE HOME & NEW AGE/ WATER & CARBONATED SOFT DRINK \\

 \hline
\end{tabular}
\caption{Customer Profiles}
\label{tab:cp}
\end{table}

\section{Schematic representation}
\label{sec:sr}

\begin{figure}
    
  \centering
    \includegraphics[width=0.8\textwidth]{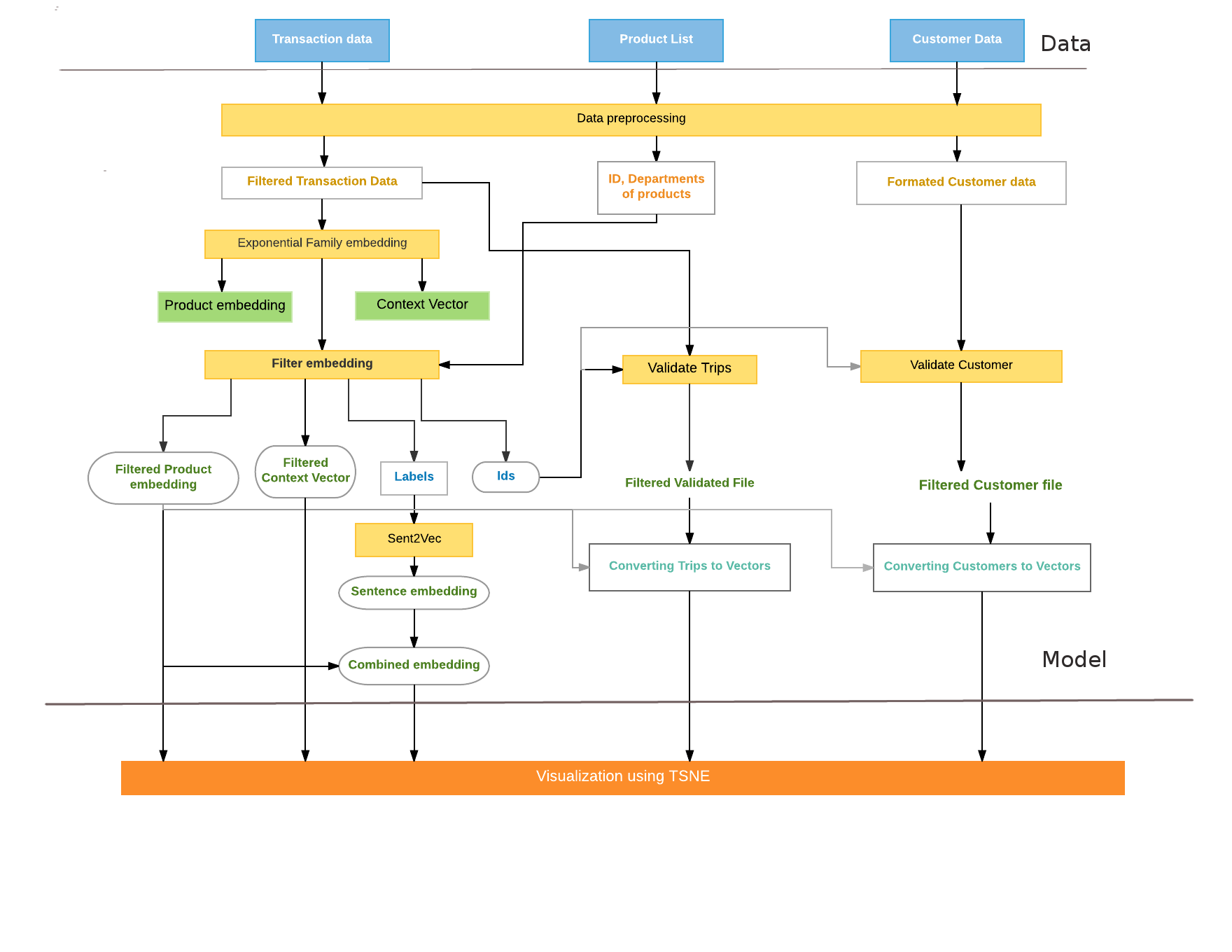}
    \caption{Pipeline for extraction of embeddings and their visualisation}
    \label{fig:dvrps}
    
\end{figure}

Figure \ref{fig:dvrps} shows the semantic data model, which includes the abstracted information of the embedding module. The module contains mainly two layers, which include Data and Model. The data layer contains the instances and features for the further processing. On top of the Model layer data go through the preprocessing stage.

The model layer converts the features in to vectors representations by using exponential family embedding as the tool. These vector representations, as discussed earlier are embeddings for products, trips and customers. The basic two vectors are product embedding and context vectors. After performing the filter embedding all the embedding will be in N dimensions, which is converted to M-dimension (2D) before visualization. 
The visualization can be consider as the final layer of the module, which is responsible for the graphical representation of the all embeddings by the help of required plugins.

\section{Conclusion}
\label{sec:con}
We have proposed a recommendation system with a novel approach that uses product embeddings, trip embeddings and customer embeddings to recommend products. We begin with Ef-emb to generate product embeddings and recommend products that are similar or co-occurring. We also created combined embeddings by combining product embeddings with sentence embeddings to bring linguistic patterns into shopping patterns. Then we use product embedding to generate trip and customer embeddings. Basically, we tried generalising the patterns at the product level to the higher understanding at the trip(shopping cart) level and customer level. We also demonstrate visually the presence of meaningful clusters in all embeddings and propose ways to recommend products by doing cluster analysis.

The recommendation of combined embeddings model needs to be validated mathematically whether they are better in terms of quality when compared to recommendations using product embeddings and sentence embeddings. Currently, we are evaluating the quality manually. 

We have proposed a recommendation system based on customer, trips and products. But we have not proposed techniques to validate them. The problem with recommendation systems is that they have online verification algorithms \cite{gori2007itemrank} \textit{i.e.} if a user clicks on one of the recommendations then the score is given inversely proportional to the rank of that recommendation which was clicked. We need to build a front end where user clicks can be recorded and our recommendation system gets a score.

\bibliography{mybib}{}
\bibliographystyle{plain}

\end{document}